\documentclass{ws-ijmpa}
\usepackage{amssymb,graphicx,amsmath,graphics,xfrac,url,subfigure}
\usepackage[super,compress]{cite}
\usepackage{multirow}
\begin{document}

\title{Study of the Klein--Gordon equation for a hydrogenic model of dyons}

\author{Edison Fernando García Veloz}
\address{School of Physical Sciences and Nanotechnology, Yachay TECH University, 100119-Urcuqu\'i, Ecuador.\\
edison.garcia@yachaytech.edu.ec}

\author{Clara Rojas}
\address{School of Physical Sciences and Nanotechnology, Yachay TECH University, 100119-Urcuqu\'i, Ecuador.\\
crojas@yachaytech.edu.ec}

\maketitle

\pub{Received (\today)}{Revised (Day Month Year)}

\begin{abstract}
This article presents the generalization of a zero spin hydrogen atom to a relativistic atomic model of hydrogen with dyons using the Klein--Gordon equation. The derivation of the Klein--Gordon equation for the particle of relative motion is shown. In addition, the analytical solutions of the equation are calculated in terms of Whittaker functions and Jacobi weighted polynomials. The discrete spectrum of energy, and the charge density of the orbiting dyon are presented. For a system of positive magnetic and electric charges in the nucleus and negative charges for the orbiting particle, and considering the first allowed values of $N$ and $l$, it was found that the dyon atom acts with a greater force of interaction between the charges of the nucleus and the secondary particle compared to the standard atom. It was obtained by comparing the distance between the nucleus and charge density concentrations from the dyon atom  with the relativistic pionic atom.

\end{abstract}

\ccode{PACS Nos.: 03.65.Pm, 03.50.De, 14.80.Hv}

\section{Introduction}	
Bound states have been used to explore several systems such as the hydrogen atom under relativistic and non-relativistic theoretical frameworks. In such systems, energy, probability densities, charge density, as well as angular momentum and other physical magnitudes are calculated and analyzed \cite{WG1,QM1}. In addition, other atomic systems have been studied in Nature. Specifically, exotic atoms are atoms in which one of the electrons has been replaced by a negatively heavy charged particle such as muon, K-meson, pion, $\Sigma$-hyperon, etc\cite{exotic}. In this research area, the description of phenomena at high energies requires the investigation of relativistic wave equations. In particular, the Klein--Gordon equation correctly describes spinless relativistic particles like pions.

In  1969, Julian Schwinger proposed a new kind of particle with both electric and magnetic charge without violating any of the results given by Dirac \cite{Sch,Dirac}. This dual charged  particle is called dyon, and several features of this particle have been theoretically researched. In particular, the interaction between a dyon and other charged particles in unbound states such as scattering can be found in the literature\cite{SchS,ScaBorn,Rp}.

Some efforts to find theoretical and experimental evidence for dyons can be found in grand unified theories which include the $SU(N)$ \cite{su}, or string theory\cite{string}. In addition, recent searches of dyons includes the use of the Full MoEDAL Trapping Detector via the Drell–Yan mechanism in $13$ TeV pp collisions. However, no candidates have been found \cite{exp}. Furthermore, some limits in the search of magnetic monopoles have been found. In particular, the Fermilab obtained a cross section limits of $2\times10^{-34}$ cm$^{2}$ for monopole masses below $850$ GeV, and the CDF experiment found a  production cross section limit for spin$-1/2$ monopoles below $0.2$ pb for masses between 200 and $700$ GeV. Also, monopole--anti-monopole bound states  have been studied under  $SU(5)$ theories considering a reduced mass of $10^{16}$ GeV to get general properties. \cite{monopol}

Dyon systems  have been investigated about bound states with spin and using a non-relativistic theoretical frameworks\cite{Bound1,Y1}, as well as bound states using magnetic monopoles \cite{Bound2,Jop}. The generalization of the spinless hydrogen atom to spinless relativistic hydrogenic atom is presented in this project by replacing both proton ($+e_{0}$) and the electron ($-e_{0}$) with dyons of charges ($e_{1},g_{1}$), ($e_{2},g_{2}$), respectively.

This article is structured in the following way. In Section 2 we present the Klein--Gordon equation for the dyon of relative motion. Section 3 shows the analytical solutions, both the radial and the angular solutions in terms of special functions. The dyon energy is discussed in section 4 . Section 5 is devoted to the charge density, and finally in section 6 we present our conclusions.
 
%%%%%%%%%%%%%%%%
\section{Klein--Gordon Equation for the  Particle of Relative Motion}
%%%%%%%%%%%%%%%%

In the relativistic quantum field theory, for dyons there exist two theories characterized by the distinct duality--symmetry groups SO$(2)$ and $Z_{4}$\cite{z4}, which relate the charges of dyons as $e_{r}g_{s}-e_{s}g_{r}=4\pi n_{rs}$ and  $e_{r}g_{s}=2 \pi n_{rs}$ for all integer numbers $n_{rs}$, respectively. These relations imply the quantization condition of dyons at non-relativistic domain : $e_{2}g_{1}-e_{1}g_{2}=2\pi n$\cite{schmin}, where $n$ is integer. These conditions are generalizations of the original Dirac's  quantization condition for electric charges and magnetic monopoles: $eg=2\pi n$ \cite{R,monR}.

On the other hand, the classical dynamics of two dyons with arbitrary electric and magnetic charges ($e_{1},g_{1}$) and ($e_{2},g_{2}$) can be expressed in terms of two decoupled motion equations: center of mass equation and relative equation. In particular, the Lagrangian for the  particle of relative motion is given by \cite{C1}:
\begin{equation}
\mathcal{ L}_{relative}=T- U=\frac{1}{2}m\upsilon^{2}-A^{0}+\frac{{\bf v}}{c}\cdot {\bf A},
\end{equation}
where $m$ is the reduced mass, $q=e_{1}e_{2}+g_{1}g_{2}$, and $g=e_{2}g_{1}-e_{1}g_{2}$ are effective charges. Here, the scalar potential is given by $A^{0}=q/4\pi r$, and ${\bf A}$ is the vector potential such that \cite{C1}
\begin{equation}
{\bf A}=\frac{g}{4\pi}\frac{1-\cos(\theta)}{r\sin(\theta)}\hat {\phi}.
\end{equation}
Setting the inertial reference frame attached to the center of mass, then the Lagrangian for the two-body problem of dyons is
\begin{equation}
\mathcal{ L}=\mathcal{ L}_{CM}+\mathcal{ L}_{relative}=\mathcal{ L}_{relative}.
\end{equation}
In order to promote the classical system into a quantum relativistic system, the  canonical quantization is necessary. The Lagrangian (1) can be interpreted as a description of a fictitious particle with charge equal to 1 under the influence of an external electromagnetic field generated by another fictitious  particle with charge ($q,g$). Then, the quantum operators for energy and momentum are given by 
\begin{equation}
\hat{E}\rightarrow i\hbar\frac{\partial}{\partial t}-A_{0},\hspace{1cm}
\hat{{\bf p}}\rightarrow -i\hbar\nabla-\frac{{\bf A}}{c}.
\end{equation}
Then, the Klein--Gordon equation for the dyon of relative motion is obtained by substituting the four-momentum operator into the Einstein's mass--energy relation $\hat{p}^{\mu}\hat{p}_{\mu}=m^{2}c^{2}$. So, considering natural units the resulting equation is
\begin{equation}
\left(i\frac{\partial }{\partial t}-A_{0}  \right)^{2}\psi=\left[ \left(i{\bf \nabla}+ {\bf A} \right)^{2}+m^{2} \right]\psi.
\end{equation}

%%%%%%%%%%%%%%%%
\section{Radial and Angular Solutions}
%%%%%%%%%%%%%%%%

To find a stationary solution for the equation, we consider the method of separation of variables in spherical coordinates so that the wave function is  $\psi=R(r)Y(\theta,\phi)e^{-iEt} $. Therefore, 
\begin{equation}
\left[r^{2}(E-A_{0})^{2}+ \left( \frac{r^{2}}{R}R''+\frac{2r}{R} R'\right)-m^{2}r^{2}\right] =\frac{Q}{Y},
\end{equation}
where $A=|{\bf A}|$ and
\begin{equation}
Q=
Y A^2+2 i A \csc (\theta )\frac{ \partial Y}{\partial \phi}-\frac{\partial^{2} Y}{\partial \theta^{2}}
-\cot (\theta ) \frac{\partial Y}{\partial \theta}-\csc ^2(\theta )\frac{\partial^{2} Y}{\partial \phi^{2}}.
\end{equation}
So, the equations must be related by a constant $\lambda$ such that  $Q=\lambda Y$. This constant can be related to the quantum number $l$ if the angular momentum for the dyon system is taken into account. 

The classical angular momentum of two arbitrary dyons is given by the contribution from both electromagnetic fields (${\bf L}_{em}$) and particles (${\bf L}_{p}$). That is\cite{C1},
\begin{equation}
{\bf L}={\bf L}_{em}+{\bf L}_{p}=\frac{e_{1}g_{2}-e_{2}g_{1}}{4\pi }\frac{\bf r}{r} +{\bf r}\times m{\bf v}+{\bf r}_{CM}\times (m_{1}+m_{2}){\bf v}_{CM}.
\end{equation}
If the coordinate system is located in the center of mass, then the quantum case for the angular momentum can be derived using the momentum operator shown in Eq. (4). Therefore, 
\begin{equation}
{\bf \hat{ L}}={\bf \hat{r} }\times\left( -i\nabla-{\bf A}\right)+\frac{e_{1}g_{2}-e_{2}g_{1}}{4\pi}{\bf e}_{r}.
\end{equation}

The operator ${\bf L}^{2}$ is identical to the operator shown in Eq. ($7$) except for a constant $\mu^{2}=g^{2}/16\pi{^2}$. By adding the constant $\mu^{2}$ into Eq. ($7$), then the eigenvalue problem for this operator can be written as

\begin{equation}
{\bf L}^{2}Y=(\lambda+\mu^{2}) Y.
\end{equation}
In this way, the eignvalues of $\hat{{\bf L}}^{2}$ can be written as $ (\lambda+\mu^{2})  =l(l+1)$. 
\subsection{Radial solution}
To solve the radial part of ($6$), let's consider $R=\Lambda(r)/r$, and energy solutions in the range $ - m<E<m$ to get  bound states in the system. Then, the variable changes $b=2\sqrt{m^{2}-E^{2}}$, $z=br$, $\nu=\sqrt{\left(l+1/2 \right)^{2}-g^{2}/16\pi{^2}-q^{2}/16\pi^{2} }$,  $ \xi=-2qE/4\pi b $  are well defined, and they give the following differential equation
\begin{equation}
 \frac{d^{2}\Lambda}{dz^{2}}+\left[ -\frac{1}{4}+\frac{\xi}{z}-\frac{\nu^{2}-1/4}{z^{2}} \right]\Lambda=0.
\end{equation}
The general solution for the Whittaker's equation ($11$) is expressed in terms of the linear independent functions $M_{\xi,\nu}(z)$ and  $M_{\xi,-\nu}(z)$ \cite{MAG}, where 

\[
M_{\xi,\nu}(z)=e^{-z/2}z^{(1/2)+\nu} F_{1}(\nu+1/2-\xi,2\nu+1,z),
\]
\begin{equation}
M_{\xi,-\nu}(z)=e^{-z/2}z^{(1/2)-\nu} F_{1}(1/2-\nu-\xi,1-2\nu,z), \hspace{0.1cm} F_{1}(a,c,z)=\sum_{n'=0}^{\infty}\frac{(a)_{n'}}{(c)_{n'}}\frac{z^{n'}}{n'!}.
\end{equation}

By imposing normalization conditions in the asymptotic behavior and integrability conditions in the origin, then the  radial function takes the form of 
\begin{equation}
R=\frac{\Lambda}{r}=\frac{1}{r}(b r)^{\nu+1/2}e^{-br/2} F_{1}(-N,2\nu+1,br),
\end{equation}
where $N$ is the highest degree of the polynomial $F_{1}$.

The radial wave function depends on several parameters which include the effective charges $q$ and $g$. Some effects such as dilation can be observed on the radial function due to different configuration of charges. In particular, under the quantization condition of $Z_{4}$ symmetry, the dyon atom composed by the charges $(+e_{0},g_{r})$ and $(+e_{0},g_{s})$  can exhibit this behavior as can be seen in the Fig. 1.

\begin{figure}[h]
\centerline{\includegraphics[width=4.0in]{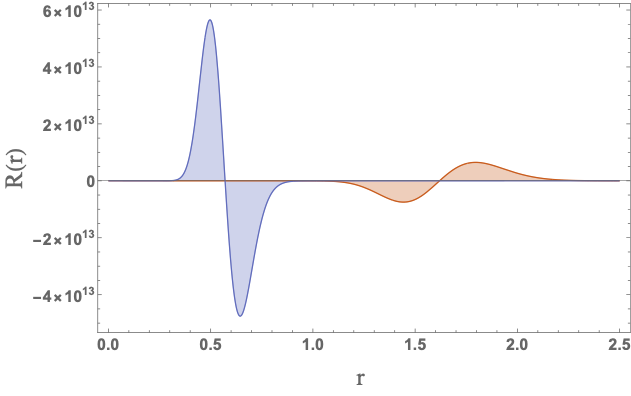}}
\vspace*{8pt}

\caption{Normalized radial functions for two dyon systems with charges related by the quantization condition  $e_{r}g_{r}=2\pi n_{rr}$ and $e_{s}g_{s}=2\pi n_{ss}$ using $e_{r}=e_{s}=+e_{0}$, and the quantum numbers $l=90$, $N=1$. The blue curve is plotted with $n_{r}=2,n_{s}=-1$, and the orange curve with $n_{r}=-1, n_{s}=-1$.\protect\label{fig4}} 
\end{figure}

\subsection{Angular solution}
Considering the angular momentum operator Eq. ($9$), then the operators ${\bf L}^{2},L_{3}$ obey the standard commutation relation $ [{\bf L}^{2},L_{3}]=0$. Since ${\bf L}^{2}$ commutes with $L_{3}$, it is possible to find a common basis of eigenfunctions for the two operators. Let be $Y(\theta,\phi)$ the common eigenvector, then 
\begin{align}
&L_{3}Y=\left(-i\frac{\partial}{\partial \phi}- \mu\right)Y=KY,\\
&Y=\Theta(\theta)e^{i(\mu+K)\phi}=\Theta(\theta)e^{ik\phi}, \hspace{1cm} k=K+\mu.
\end{align}
The eigenvalue $K$ and the quantum number $l$ must satisfy the constraints $l=|\mu|,|\mu|+1,|\mu|+2,...,$and$ -l\leq K\leq l$ simultaneously \cite{AM}.

The angular equation ($7$) and the angular wave function $Y$ with the variable changes $x=\cos(\theta)$ ,$\Theta=(1-x)^{-(a+b)/2}(1+x)^{-(b-a)/2}\omega$ lead to the differential equation
\begin{equation}
\left(x^2-1\right) \omega ''(x)-2[a+(b-1) x] \omega '(x)+ (-b-\lambda ) \omega (x)=0,
\end{equation}
where $a=k-g/4\pi $ and $ b=g/4\pi$.  The additional variable changes $z=(1+x)/2$ , $\alpha+\beta+1=2(1-b), \alpha \beta=-b-\lambda,\gamma=a-b+1$ give the hypergeometric differential equation
 \begin{equation}
 z (1-z) \frac{d^{2} \omega}{dz^{2}}+ [\gamma -(\alpha+\beta+1)z] \frac{d \omega}{dz}-\alpha \beta \omega=0.
\end{equation}

The general solution of Eq. ($17$)  is given in terms of hypergeometric functions $F$ of second order
\begin{equation}
\omega(z)=C_{1}F(\alpha,\beta,\gamma,z)+C_{2}z^{-\gamma+1}F(\alpha-\gamma+1,\beta-\gamma+1,2-\gamma,z),
\end{equation}
where 
\begin{equation}
F(\alpha,\beta,\gamma,z)=\sum_{n=0}^{\infty}\frac{(\alpha)_{n}(\beta)_{n}}{(\gamma)_{n}}\frac{z^{n}}{n!}.
\end{equation}

The function $F$ is finite if it is reduced to a finite polynomial \cite{MAG}. Then to get physically acceptable solutions, the series must terminate at a certain power $n=0,1,2,...$ The condition is achieved, if $\alpha$ or $\beta$ is a negative integer. Let be $\alpha=-n$ and returnig some variable changes, therefore 
\begin{equation}
\alpha+\beta+1=2(1-b),
\end{equation}
\begin{equation}
l(l+1)=n^{2}-\mu+n-2n\mu+\mu^{2}=(n-\mu)(n-\mu+1).
\end{equation}

The relation ($21$) shows that $l=n-\mu$.  Considering the variables changes  $-2b=\hat{\alpha}+\hat{\beta}, 2a=\hat{\beta}- \hat{\alpha}$ in Eq. ($16$), then 
\begin{equation}
(1-x^{2})\omega''+[\hat{\beta}-\hat{\alpha}-(\hat{\alpha}+\hat{\beta}+2)x]\omega'+n(\hat{\alpha}+\hat{\beta}+n+1)\omega=0,
\end{equation}
which is satisfied by the Jacobi polynomials    $\omega=P_{n}^{\hat{\alpha},\hat{\beta}}(x)$. So, the angular function is such that
\begin{equation}
Y_{\mu l K}(\theta,\phi)=(1-x)^{-(a+\mu)/2}(1+x)^{-(\mu-a)/2}P_{n}^{\hat{\alpha},\hat{\beta}}(x) e^{i(K+\mu )\phi}.
\end{equation}

The squared modulus of the the generalized spherical harmonics $(23)$ can be affected by the system of charges. Specifically, the intensity of the modulus could decrease under an increase of magnetic charges, as can be seen in  Figs. 2 and 3. 

\begin{figure}[h]
\centerline{\includegraphics[width=4.0in]{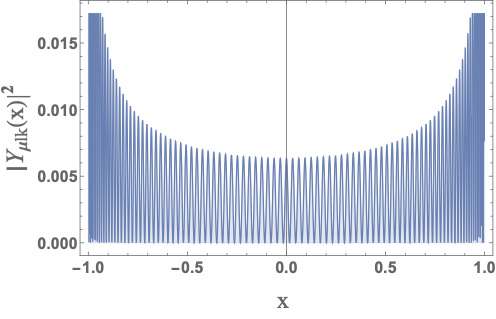}}
\vspace*{8pt}
\caption{Squared modulus for a dyon system with charges related by the quantization condition  $e_{r}g_{r}=2\pi n_{rr}$ and $e_{s}g_{s}=2\pi n_{ss}$ using $e_{r}=e_{s}=+e_{0}$, and the quantum numbers $l=100$, $K=0$. The curve is plotted with $n_{r}=-1,n_{s}=-1$.\protect\label{fig4}} 
\end{figure}

\begin{figure}[h]
\centerline{\includegraphics[width=4.0in]{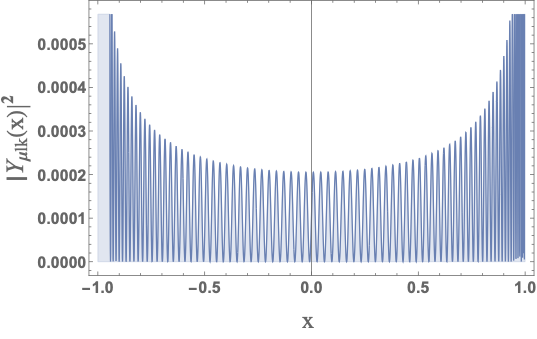}}
\vspace*{8pt}

\caption{
Squared modulus for a dyon system with charges related by the quantization condition  $e_{r}g_{r}=2\pi n_{rr}$ and $e_{s}g_{s}=2\pi n_{ss}$ using $e_{r}=e_{s}=+e_{0}$, and the quantum numbers $l=100$, $K=0$. The curve is plotted with $n_{r}=-6,n_{s}=-1$.\protect\label{fig4}} 
\end{figure}
\section{Energy}
Let be $m_{1}$, $m_{2}$ the mass of the dyon nucleus and the mass of the dyon under the influence of $m_{1}$, respectively. The masses relation $m_{1}\gg m_{2}$ for the hydrogen like atom implies  $m\approx m_{2}$. Also, the center of mass reference frame is located approximately near the mass $m_{1}$ because of $m_{1}\gg m_{2}$. Taking into account these approximations, the wave function $\psi$ corresponds to the dyon with $m_{2}$. Then, the energy for this dyon is reached by noting that the normalization condition in the solutions $(12)$ indicates the following relation.
\begin{equation}
\nu+1/2-\xi=-N.
\end{equation}
By clearing algebraically the term of the energy, then the resulting expression is
\begin{equation}
E=E_{Nl\mu}=\pm m\left[1-\frac{q^{2}}{q^{2}+16\pi^{2} \left(\sqrt{ (l+1/2)^{2}-g^{2}/16\pi{^2}-q^{2}/16\pi^{2}}+1/2+N\right)^{2}} \right]^{1/2}.
\end{equation}

Let's consider the monopolonium case, that is, the monopole--anti-monopole bound state. By setting the electric charges equal to zero $e_{1}=e_{2}=0$, $g_{1}=Zg_{0}$, $g_{2}=-g_{0}$, and considering the changes $e_{1,2}\rightarrow\sqrt{4\pi}e_{1,2}$,  
$g_{1,2}\rightarrow \sqrt{4\pi}g_{1,2}$ to remove the Lorentz-Heaviside units, then the energy $(25)$ is reduced to

\begin{equation}
E=- m\left[1-\frac{(Zg_{0}^{2})^{2}}{(Zg_{0}^{2})^{2}+\left(\sqrt{ (l+1/2)^{2}-(Zg^{2}_{0})^{2}}+n-l-1/2\right)^{2}} \right]^{1/2}.
\end{equation}
where $n=N+l+1$ is the principal quantum number, and the elementary charges are related by the original Dirac's quantization condition $e_{0}g_{0}=1/2$.

Expanding the relation (26) in a series of powers of $Z$, then
\begin{equation}
E=m-\frac{m(Zg_{0}^{2})^{2}}{2n^{2}}+O(Z^{4})
\end{equation}

The quadratic term in (27) can be identified as the non-relativitic energy for the monopolonium. This relation is in agreement with the results obtained in the literature for $Z=1$ \cite{groupmono,monopol}. For the relativistic case, the energy (26) also depends on the quantum number $l$. This feature imposes a restriction on the allowed energies of the system. 
In particular, considering a reduced mass of $m=10^{16}$GeV, $Z=1$ and and the minimun allowed quantum number $l=34$ for this system, some values of the binding energy $E_{b}=E-m$ for the monopolonium  are shown in the table 1.  These values differ to a larger extent in the extreme values and slightly for the intermediate quantities 
reported by Hill\cite{monopol} for the non-relativistic monopolonium, as can be seen in the right-hand side of the table 1.

\begin{table}
\begin{center}
\begin{tabular}{ |p{3cm}|p{3cm}|p{3cm}| }
 \hline
 \multicolumn{3}{|c|}{ Binding Energy (GeV)} \\
 \hline
 Principal quantum number $n$ & Relativistic regime with $l=34$&Non-relativistic regime \\
 \hline
 $4.17\times 10^{1}$   & $6.85\times 10^{15}$    &$3.35\times 10^{15}$\\
 $4.17\times 10^{2}$&  $3.90\times10^{13}$  & $3.35\times 10^{13}$\\
 $4.17\times 10^{3}$ &$3.42\times10^{11}$ &$ 3.35\times 10^{11}$\\
 $4.17\times 10^{4}$ &$3.37\times 10^{9}$ & $3.35\times 10^{9}$\\
 $4.17\times 10^{5}$&  $3.37\times 10^{7}$  & $3.35\times 10^{7}$\\
 $4.17\times 10^{6}$& $3.37\times 10^{5}$  & $3.35\times 10^{5}$\\
 $4.17\times 10^{7}$& $3.37\times 10^{3}$  & $3.35\times 10^{3}$\\
 $4.17\times 10^{8}$& $34$  & $3.35\times 10^{1}$\\
 $4.17\times 10^{9}$& $2$  & $3.35\times 10^{-1}$\\
 \hline
\end{tabular}
\caption{\label{tab:table-name} Binding energies for the Monopolonium.}
\end{center}
\end{table}

On the other hand, by setting the magnetic charges equal to zero $g_{1}=g_{2}=0$ in the energy expression (25),  it is possible to  reproduce the expected energy for the hydrogen atom. So that, $q^{2}=(-Ze_{0}^{2})^{2}=(Z\alpha 4\pi)^{2}$, where $\alpha$ is the fine structure constant and $Z$ the number of electric charges in the nucleus. That is,

\begin{equation}
E=E_{Nl\mu}=- m\left[1-\frac{(Z\alpha 4\pi)^{2}}{(Z\alpha 4\pi)^{2}+16\pi^{2} \left(\sqrt{ (l+1/2)^{2}-(Z\alpha 4\pi)^{2}/16\pi^{2}}+1/2+N\right)^{2}} \right]^{1/2}.
\end{equation}

\begin{equation}
E=E_{Nl}=- m\left[1+\frac{(Z\alpha)^{2}}{\left(\sqrt{ (l+1/2)^{2}-(Z\alpha)^{2}}+1/2+N\right)^{2}} \right]^{-1/2}.
\end{equation}

In order to find the allowed values for the magnetic and electric charges in the hydrogen atom with dyons, the quantization conditions must be considered. Here, we consider the constraint between the charges under the $Z_{4}$ symmetry. So, for the same dyon the condition $e_{r}g_{r}=2\pi n_{rr}$ with $n_{rr}=1$ implies that the value of the elementary magnetic charge is $g_{0}=\pm 2\pi/e_{0}$. 

To illustrate the behaviour of the energy Eq. (25) under an increasing number of electric and magnetic charges in the nucleus, a particular case in which there are $Z$ charges in the nucleus is considered. For this system with a point nucleus that contains $Z$ elementary charges $(e_{1},g_{1})=(Z e_{0}, Z 2\pi/e_{0})$, and the second particle is such that $(e_{2},g_{2})=(-e_{0},-2 \pi/e_{0})$, then the effective charge is $g= 0$, and satisfies the previous quantization conditions for dyons. Some energy curves are shown in Fig. 4.

\begin{figure}[th!]
\centering
\includegraphics[width=\textwidth]{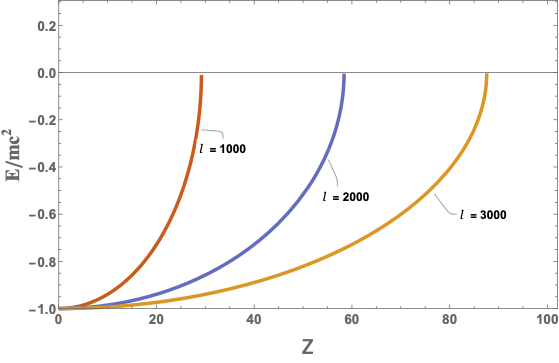}
\caption{Energy curves for different $l$ values and fixed $N=0$ for the dyon--dyon system and curves for fixed $l=2000$}
\label{fig:1a}
\end{figure}

\begin{figure}[th!]
\includegraphics[width=\textwidth]{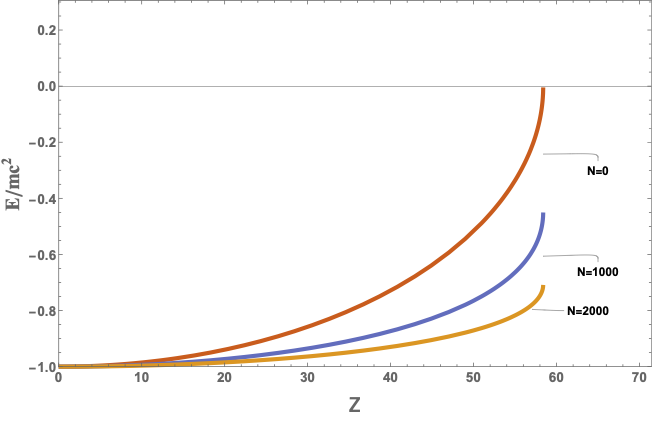}
\caption{Energy curves for different $l$ values and fixed $N=0$ for the dyon--dyon system  and several $N$ values (b).\protect\label{fig1}}
\label{fig:1b}
\end{figure}

This system can not show a transition from positive energy to a negative one under an increasing nuclear charges as can be noticed in the limit $Z\rightarrow \infty$ in the energy expression Eq. $(25)$.

For the system of charges with $Z=1$, the allowed energies for the system start from $l=34$ because of the square root in the energy expression. If we introduce the principal quantum number $ n_{p}=N+l+1$, then the first levels $1s, 2p,3s,3p ,$etc. are forbidden states for the dyon--dyon system. In addition, since $\mu=0$ the energy is reduced to $E_{n_{p}l\mu}=E_{n_{p}l}$. So that for a  level $l$ and a given $n_{p}$, the system  has different energies. However, the $2l+1$ degeneracy related with the quantum number $K$ remains. 

\section{Charge Density}
To find the charge density expression, the Klein--Gordon equation $(5)$ is multiplied by $\psi^{*}$ from the left-hand side and subtracted the complex conjugate. So that
\begin{equation}
0=\eta^{\mu \nu}\left[\frac{\partial}{\partial x^{\mu}}\left( \psi\frac{ \partial}{\partial x^{\nu}}\psi^{*}- \psi^{*}\frac{ \partial}{\partial x^{\nu}}\psi\right) -2\frac{\partial}{\partial x^{\mu}}\left(\psi iA_{\nu}\psi^{*}   \right) \right].
\end{equation}

Denoting $\eta^{\mu \nu}$ as the Minkowski metric component, and  $J_{\nu}$ as 
\[
J_{\nu}=\frac{i e}{2m}\left(\psi^{*}\frac{\partial}{\partial x^{\nu}}\psi-\psi\frac{\partial}{\partial x^{\nu}}\psi^{*}  \right)-\frac{e}{m}A_{\nu}\psi \psi^{*},
\]
then the relation $(30)$ shows the continuity equation $\partial_{\mu} J^{\mu}=0$. The term $J_{\nu}$ is interpreted as the four-current density of electric charge.  Since the dyon carries both magnetic and electric charges at the same time, then the four-current associated to the magnetic charge is the same but with the replacement $e_{0} \rightarrow g_{0}$.

Taking  the wave function for the stationary state $\psi=R(r)Y(\theta,\phi)e^{-iEt}=\psi({\bf r}) e^{-iEt}$, then the electric charge density and the magnetic charge density are given by the expressions 

\begin{equation}
\mathcal{P}_{e} =- e_{0}\left(\frac{E-A_{0}}{m} \right)\psi({\bf r})\psi({\bf r}),\hspace{1cm} \mathcal{P}_{g} =- g_{0}\left(\frac{E-A_{0}}{m} \right)\psi({\bf r})\psi({\bf r}).
\end{equation}

To explore the charge density for the system with $Z=1$, let's consider the radial charge density per unit charge $\mathcal{P}_{r}/(e,g)$= $(E-A_{0})R^{2}r^{2}/m$. By noting that the density function can be written as $\mathcal{P }_{r}/(e,g)=\mathcal{P}_{r}(br)$, and $0<b<1$ if $l$ or $N$ increases, then the effect on the density function is a horizontal dilation. That is, the charge density moves to regions further away from the nucleus. 

Figs. 5 and 6 show the curves of $\mathcal{P}_{r}/(e,g)$ for a system with mass $m=139.577$ MeV. The plots indicates that the charge density concentration moves away from the nucleus when $N$ increases, but the density is gradually distributed to regions surrounding the nucleus. Similarly when $l$ increases, the concentration shifts to the right. This is the same behavior than the relativistic pionic atom using the Klein--Gordon equation, as can be seen in Figs. 7 and 8.

Furthermore, Figs. 5, 6, 7 and 8 indicate that the charge density concentrations for the initial values of $N$ or $l$ are closer to the nucleus for the dyon, and for the relativistic pionic atom are far away from the nucleus. Therefore, in these levels, it indicates that this particular dyon system acts like a pionic atom or a standard hydrogen atom with a stronger force of attraction between the proton and a negative pion.

By setting the magnetic charges equal to zero, $\mathcal{P}_{g}=0$, ${\bf A}={\bf 0}$, $A_0=q/4\pi r=-Ze^{2}_{0}/4\pi r=-Z\alpha/r$. Then, the radial electric charge density for the relativistic hydrogen atom with zero spin is reproduced. That is, 

\[
\mathcal{P}_{r}= e_{0}(E+Ze^{2}_{0}/4\pi r)R^{2}r^{2}/m
\]
where the radial function is such that
\[
R=\frac{1}{r}(b r)^{\nu+1/2}e^{-br/2} F_{1}(-N,2\nu+1,br), \hspace{1cm} \nu=\sqrt{ (l+1/2)^{2}-(Z\alpha)^{2}}.
\]
\begin{figure}[h]
\centerline{\includegraphics[width=4.0in]{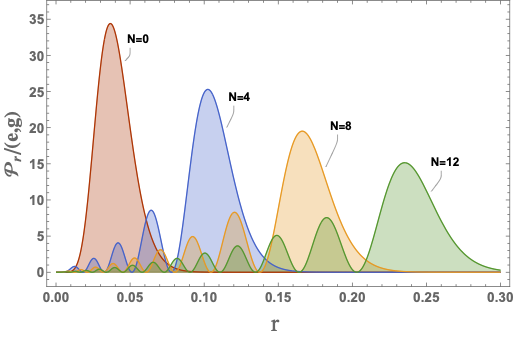}}
\vspace*{8pt}
\caption{Curves of $\mathcal{P}_{r}/(e,g)$ for $l=34$  and  $N=0, N=4, N=8,  N=12.$\protect\label{fig2}}
\end{figure}

\begin{figure}[h]
\centerline{\includegraphics[width=4.0in]{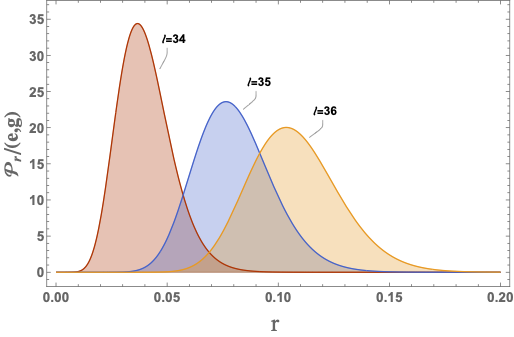}}
\vspace*{8pt}
\caption{Curves of $\mathcal{P}_{r}/(e,g)$ for $ N=0$ and $l=34, l=35,  l=36.$\protect\label{fig3}}
\end{figure}

\begin{figure}[h]
\centerline{\includegraphics[width=4.0in]{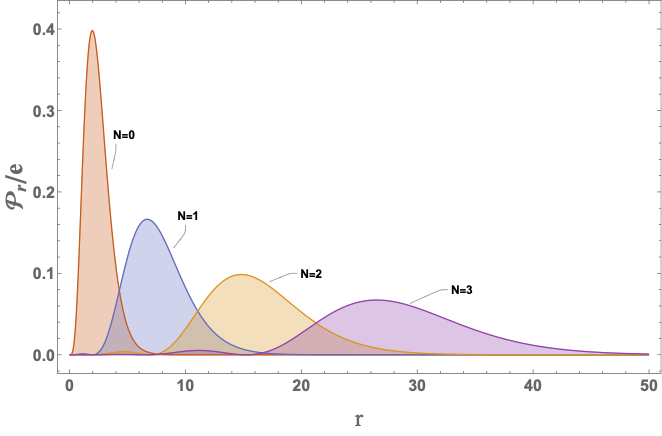}}
\vspace*{8pt}
\caption{Curves of $\mathcal{P}_{r}/e$ for the pionic atom with $ l=0$ and $N=0,  N=1, N=2,N=3$.\protect\label{fig4}} 
\end{figure}

\begin{figure}[h]
\centerline{\includegraphics[width=4.0in]{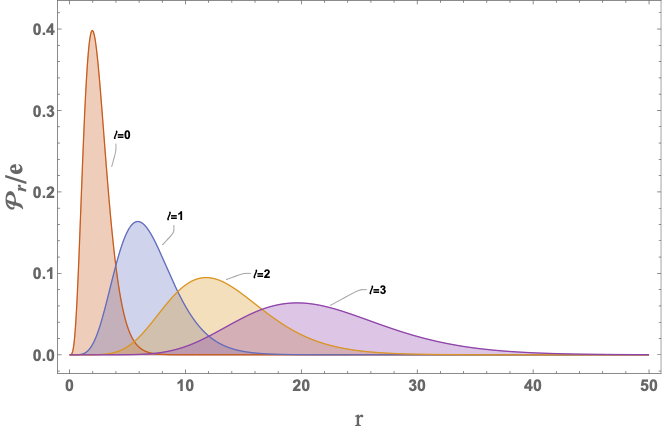}}
\vspace*{8pt}
\caption{Curves of $\mathcal{P}_{r}/e$ for the pionic atom with $ N=0$ and $l=0,  l=1, l=2,l=3$\protect\label{fig5}}
\end{figure}

\section{Conclusions}
In this article, we have shown a study of an hydrogenic atomic model composed by two dyons. We have used the classical description of a two-body system of dyons under the Lagrangian formalism, and the canonical quantization into the Einstein mass--energy relation to promote the system to a relativistic quantum approach.  From the resulting Klein--Gordon equation, it was considered to find solutions for stationary states of the system. 

In addition, it is shown the discrete energy depends on three quantum numbers: $N, l, \mu$. For a system with a nucleus of variable electric and magnetic charge, it did not show a transition of energy sign  under an increasing of $Z$ charges in the nucleus. Furthermore, if $Z=1$, the first levels $ 1s, \,2p, \,3s, \,3p, $ etc.  are forbidden states, and the degeneracy of the energy exhibits a value of $2 l+1$. It was also found that the dyon atom with $Z=1$ and $m=139.577$ MeV concentrates the radial charge density in regions closer to the nucleus than the  pionic atom under the Klein--Gordon equation. Also, by setting the magnetic charges equal to zero, the radial charge density and energy expressions for the standard hydrogen atom are restored, as well as the monopolonium case was generated.

In this article, the nucleus is considered as a point particle. However, hydrogen-like atoms contain a nucleus’s finite extent. Then, this suggests improving  the model of the dyon atom considering the size of the nucleus using  a modified scalar or vector potential. This consideration could show bound states for larger $Z$ charges  and a smaller allowed quantum number $l$  than the values found in this work.

%%%%%%%%%%%%%%%%
%\bibliographystyle{unsrt}
%\bibliography{sample.bib}
%%%%%%%%%%%%%%%%%
%\end{document} 

%%%%%%%%%%%%%%%%
\bibliographystyle{unsrt}

\end{document}